# Developing controlled natural language for formal specification patterns using AI assistants

Natalia Garanina
Cyber-Physical Systems lab
Institute of Automation and Electrometry
Novosibirsk, Russia
natta.garanina@gmail.com

Vladimir Zyubin
Cyber-Physical Systems lab
Institute of Automation and Electrometry
Novosibirsk, Russia
zyubin@iae.nsk.su

Igor Anureev
Cyber-Physical Systems lab
Institute of Automation and Electrometry
Novosibirsk, Russia
anureev@gmail.com

*Abstract* — Using an AI assistant, we developed a method for systematically constructing controlled natural language for requirements based on formal specification patterns containing logical attributes. The method involves three stages: 1) compiling a generalized natural language requirement pattern that utilizes all attributes of the formal specification template; 2) generating, using the AI assistant, a corpus of natural language requirement patterns, reduced by partially evaluating attributes (the developed prompt utilizes the generalized template, attribute definitions, and specific formal semantics of the requirement patterns); and 3) formalizing the syntax of the controlled natural language based on an analysis of the grammatical structure of the resulting patterns. The method has been tested for event-driven temporal requirements.

## I. INTRODUCTION

Developing a controlled natural language (CNL) for specifying event-temporal requirements is a pressing research challenge in software requirements engineering. In today's world, where computer systems are becoming increasingly important in industries ranging from finance and healthcare to transportation and industrial automation, ensuring accurate and understandable specification of system events and their temporal characteristics plays a critical role. According to [1], the majority of errors in software development occur during the requirements specification stage, as engineers, analysts, and other stakeholders misunderstand the requirements. Any error in the initial requirements means the need for subsequent correction, which takes time and requires additional costs [2,3].

We developed a method for systematically constructing controlled natural language requirements using an AI assistant for specifying requirements based on formal specification patterns containing logical attributes. The method includes three stages: 1) composing a generalized requirement pattern in natural language that utilizes all attributes of the formal specification pattern; 2) generating, using an AI assistant, a corpus of requirement patterns in natural language, reduced by partially evaluating attributes (the developed prompt uses the generalized pattern, evaluating attributes, and specific formal semantics of requirement patterns); 3) formalizing the syntax of the resulting controlled natural language based on the analysis of the grammatical structure of the obtained patterns.

We tested the proposed method on the event-temporal requirements language EDTL [4].

Our method is suitable for patterns whose interpretation in natural language depends on constant attribute values that simplify the formula. The resulting language has formal semantics by design, unlike other methods where a requirements language is first developed and then its formal semantics is defined, such as EARS [5] and Rimay [6]. Requirements languages that already have formal semantics are also known, where several logical formulas are associated with natural language expressions, but such languages are rather limited and unexpressive [7, 8]. In other cases, such as SysReq [9], the requirements language is expressive, but its formal semantics is difficult to express and non-compact, and, as a result, difficult to verify.

Thus, the development of a controlled natural language can be carried out in various ways: by constructing it following formal semantics, or, conversely, by constructing it based on the needs (not always well formalized) of domain engineers, and then matching logical formulas with the resulting NL constructs. Currently, there are expressive CNLs with complex computable semantics (SysReq [9], BraceAssertion [10], PROPAS [11]), or inexpressive ones with simple semantics [7, 8, 12-15].

The requirements specification format for which we propose the construction of CNL assumes three concise notations: tabular (logical values of requirement attributes), formal (requirement semantic formula), and linguistic (natural language phrase). These notations are compact and simultaneously accessible, while ensuring sufficient expressiveness of requirements through the diversity of attribute value combinations. In addition to the EDTL we developed, the SUP requirements representation format developed by the German company BTC [16-18] possesses similar properties.

Our systematic CNL development technology offers the following advantages. A single initial NL pattern and the availability of three-valued Boolean attributes (with true, false, and variable values) enable the generation of a wide range of sentences from a compact initial representation. Furthermore, a single initial NL pattern facilitates working with AI assistants, i.e., writing the prompt pattern, and ensures a consistent grammatical structure for the resulting phrases, facilitating grammar construction. Furthermore, each resulting phrase has a formal semantics by design.

## II. PRELIMINARIES

Controlled Natural Language (CNL) is a subset of some natural language with an explicitly defined (restricted) syntax and vocabulary, so that its sentences can be unambiguously translated into some formal logic [19]. A typical CNL uses a well-defined subset of the grammar and vocabulary of the language, but adds terminology needed in the specific domain.

We consider formal behavior specifications for systems that can be described using the values of variables that change over time and define the system's states at different points in time. A formal requirements specification pattern is a set of attributes. Each attribute, and the set as a whole, has an intuitive, informal description of its purpose in the context of the system requirements specification, while their formal semantics are formulas of some logic. Since the formal semantics of each attribute is a logical formula, the logical values of these attributes can be true, false, or var — in general, a variable value dependent on the system's state and described by a formula over the system's variables. Below we provide examples of the EDTL and SUP formal specification patterns.

### A. Event-Driven Temporal Logic (EDTL)

***Pattern as a Set of Attributes.*** Event-Driven Temporal Logic (EDTL) is a method for formally describing requirements. This approach to recording requirements enables users to describe the behavior of control systems in terms of events and interactions with the control object and the environment, considering the control system and control object as a "black boxes."

An EDTL requirement R is a tuple consisting of six EDTL attributes: R = (trigger, invariant, final, delay, reaction, release). The attributes of an EDTL requirement are system events constrained by specified temporal relationships.

***Informal Description of Attributes. Example.*** Let's describe the informal meanings of EDTL attributes:

- *trigger* is the initial (trigger) event of the requirement;
- *invariant* is a condition that holds from the moment the trigger event occurs until a release or reaction occurs;
- *final* is an event that follows the trigger and starts waiting for a reaction;
- *delay* is an event that specifies the time interval within which the reaction event must occur from the final moment;
- *reaction* is the expected event of the requirement;
- *release* cancels the requirement: if this event occurs, the requirement is considered fulfilled, regardless of the occurrence of events of other requirement attributes.

An example of an EDTL requirement for a hand dryer:

Identify applicable funding agency here. If none, delete this text box.

*If hands are present under the dryer and the dryer is turned on, the dryer will continue to operate.* (1)

In this requirement, the trigger is the presence of hands under the dryer and the dryer is running. The response to the trigger is the dryer continuing to operate. This requirement must always be satisfied, so the release event never occurs. There are no special conditions for the dryer to continue operating after the trigger, nor are there any timing constraints, so the invariant, final event, and delay do not require specification.

***Formal semantics of the pattern and its attributes.*** The formal semantics of the EDTL pattern is the linear temporal logic formula LTL[20]:

$$\mathbf{G}\ (trig \rightarrow ((inv \land (\neg\ fin))\ \mathbf{W}\ (rel \lor fin \land$$
$$((inv \land (\neg\ del))\ \mathbf{W}\ (rel \lor (inv \land rea))))))$$

In this formula, the propositional statements *trig*, *inv*, *fin*, *rel*, *del*, and *rea* specify the logical values of the requirement attributes and are Boolean constants or propositional formulas depending on the values of the system variables.

For requirement (1), the attribute values are presented in Table 1:

TABLE I. ATTRIBUTE VALUES FOR EDTL REQUIREMENT (1)

| trigger | release | final | delay | invariant | reaction |
|---------|---------|-------|-------|-----------|----------|
| $H \land D$ | false | true | true | true | D |

Here, the Boolean variable *H* specifies the presence of hands under the dryer, and *D* specifies whether the dryer is turned on.

We have shown that the EDTL pattern satisfies the criteria for formal specification patterns to which our method for constructing controlled natural language can be applied.

### B. Simplified Universal SUP Pattern

***Pattern as a Set of Attributes.*** The Simplified Universal Pattern (SUP) from BTC is a single pattern with 15 parameters. Most parameters have default values. The BTC Embedded platform uses a graphical editor for formalization, allowing the engineer to fill in only the required parameters; parameters with default values are hidden from the user. The SUP defines the relationship between a trigger and an action. Typically, a trigger is some behavior that the system being specified can observe at its ports, and an action is some behavior it must perform in response. There are three interpretations of the SUP: progress (the trigger is followed by the action), invariant (the trigger and action occur simultaneously), and order (the action is always preceded by the trigger). Here, we use progress, the most common interpretation, from the perspective of an observer. An observer is a component that runs in parallel with the system, monitors ports, and generates an error signal if the specified behavior is violated.

*Informal Description of Attributes. Example.* The attributes of the SUP pattern are informally described as follows:

**Trigger Phase.** The trigger phase is defined by the Trigger Start Event (TSE), Trigger Condition (TC), Trigger End Event (TEE), and Trigger Exit Condition (TEC), as well as two time constraints: Trigger Min (Min) and Trigger Max (TMax). Technically, events are similar to conditions. The term "event" is used to emphasize that the condition marks the beginning or end of a phase. A SUP trigger is considered successfully captured if TEE occurs within [TMin, TMax] of TSE, and TC is satisfied within this interval. If TEC is satisfied, the observation cycle is terminated.

**Action Phase.** The Action Phase is defined by the Action Start Event (ASE), Action Condition (AC), Action End Event (AEE), Action Exit Condition (AEC), Action Min (AMin), and Action Max (AMax) parameters, similar to the Start Phase. The AEE must occur within [AMin, AMax] after the ASE, and the AC must occur in between. Otherwise, observation fails. If the AEC occurs, the current observation cycle is aborted and a new one begins. If the AEE is observed successfully, the current observation cycle succeeds and a new one begins.

**Local Response Region.** The local response region [LMin, LMax], consisting of two parameters, limits the time interval between the TEE and the ASE; if the ASE is observed too early or too late, observation fails.

The SUP monitor is a state machine, so only the first trigger firing is recorded in each observation cycle. By default, TSE = TC = TEE, ASE = AC = AEE, TEC = AEC = false, AMin = AMax = TMin = TMax = LMin = Lmax = 0. In the following example, we assume the default values for the SUP and specify only the required values.

There are two examples of requirements in text form and using the SUP (Table 2):

(A1) $inp_1$ is true once every 35 ms.

(A2) $inp_1$ is true once every 30-40 ms.

TABLE II.  ATTRIBUTE VALUES FOR SUP REQUIREMENTS

| ID | Trig<br><br>TSE =<br>TC =<br>TEE | [Lmin, Lmax] | Action | | | [Amin, Amax] |
|----|----|----|----|----|----|----|
|    |    |    | ASE | AC | AEE |    |
| A1 | true | [0, 0] | true | $\neg inp_1$ | $inp_1$ | [35ms, 35 ms] |
| A2 | true | [0, 0] | true | $\neg inp_1$ | $inp_1$ | [30ms, 40 ms] |

*Formal semantics of the pattern and its attributes.* The authors define the semantics of the SUP pattern as a counter automaton. However, it can also be defined as a formula of LTL logic, as for the EDTL pattern, but for propositional statements involving unbounded integer variables. Furthermore, in many cases, including the example case, standard LTL is sufficient. Although defining LTL semantics for SUP is beyond the scope of this paper, in the context of constructing controlled natural language, it is important that informal definitions of this pattern imply that its attributes also take the values true or false.

Attributes that are trigger and action execution conditions or their start and end markers (TSE, TC, TEE, ASE, AC, AEE, TEC, AEC) are obviously Boolean;

Attributes that are intervals are interpreted as follows:

- The local response region [Lmin, Lmax] is a constraint on the occurrence of the response (final in EDTL terms), so the interval [0, 0] corresponds to the value false – the response must occur immediately, and the interval [0, ∞) corresponds to the value true – there is no constraint on the occurrence time of the response. All other options correspond to the time-varying value of the Boolean statement LS = (t ⩾ Lmin ∧ t ⩽ Lmax), which constrains the start time of the action.

- Minimum/maximum time for start/action [Tmin, Tmax] and [Amin, Amax] – these time intervals are interpreted similarly. For a trigger, the interval [0, 0] (false) means the trigger never occurs, while the interval [0, ∞) (true) means it may occur at any time. For an action, the interval [Amin, Amax] is analogous to the EDTL delay attribute: [0, 0] (false) means the action should be immediate, while the interval [0, ∞) (true) implies that the action's completion may be delayed indefinitely.

Therefore, our proposed method for constructing controlled natural language can also be used for the SUP pattern.

III. THE METHOD

The systematic method for constructing CNLs using formal specification patterns is as follows.

*A. Stage 1: Base Pattern*

For the formal requirement specification pattern, a base specification pattern is constructed in natural language manually. It consists of one or more natural language sentences that describe the meaning of the requirement as fully yet concisely as possible, when each of its attributes is not identically true or identically false. These sentences must contain slots for substituting the values of all pattern attributes. It is desirable that the sentences be constructed so that when the attribute meanings, expressed in natural language as names or short phrases, are substituted into the slots, the resulting constructions are grammatically correct.

The result of this stage is a natural language sentence pattern containing slots for all requirement attributes.

*B. Stage 2: Derived Patterns*

It is known that if attributes take identically true or identically false values, the logical formula corresponding to the formal semantics can be significantly simplified. Simplifying the requirement formula also simplifies its expression in natural language. At this stage, it is necessary to identify how exactly the basic specification pattern in natural language is transformed when some of its attributes are constant — identically *true* or *false*. To construct new derived sentence patterns, we use an AI assistant based on LLM as follows.

*(1) Basic Prompt.* We define a text query pattern (prompt) that provides a basic pattern for the requirement specification in natural language and specifies the values of constant attributes. Additionally, it is useful to require from AI assistent an explanation of why the final result is correct. This addition allows for identifying the direction for refining queries to the AI assistant if the result is not semantically correct enough.

The basic prompt can be used without any additions. However, the following instructions to the AI assistant may be helpful to ensure greater "linguistic naturalness" and semantic correctness of the final phrase.

*(2) Formal semantics of the requirement.* For a more precise result, it may be appropriate to specify the formal semantics of the requirement for the given values of the constant attributes in each specific query. This formal semantics is calculated by substituting the constant attributes into the semantics formula of the base pattern.

*(3) Explanation of the intuitive meaning of the constant attribute values.* These explanations guide the AI assistant on how to construct the resulting sentence if the values of the logical attributes are true or false. In these cases, the names of the constant attributes do not appear in the resulting sentence at all, but their specific meaning is still revealed. Some examples are described in the Experiments section.

The result of this step will be a list of sentence patterns in natural language containing slots for attributes. These patterns are typically shorter than the original pattern.

*C. Stage 3: CNL Grammar*

Using an AI assistant to construct a list of patterns matching a single initial pattern, the resulting patterns share a similar grammatical structure through the use of LLM. This simplifies the construction of a grammar for controlled natural language. At this stage, a grammar is constructed from a list of patterns with similar grammatical structure and attribute slots. The terminal symbols are the words of the constructed patterns, along with unique slot designations for different attributes. An example is given in the next section.

IV. EXPERIMENTS WITH EDTL

We will illustrate our approach using the example of constructing a controlled natural language for the EDTL pattern *(trigger, invariant, final, delay, reaction, release)*. In our experiments, we used Google's AI mode as the AI assistant.

*A. Stage 1: Basic Pattern for EDTL*

The basic pattern for the natural language specification of EDTL is the following phrase:

*After 'trigger', 'invariant' is valid until either 'release' or 'reaction', and 'reaction' must occur within 'delay' from 'final'.*

*B. Stage 2: Derived Patterns for EDTL*

This stage generates a set of derived shorter phrases corresponding to cases where one or more attributes of the specification pattern take constant values. To generate this set, we used Google's AI assistant. We will illustrate the steps of stage 2 using the EDTL example.

*(1) Basic Prompt.* To work with the AI assistant, we defined a basic prompt pattern, which includes the basic formal specification pattern in natural language developed in the previous step:

*Reformulate in English the following sentence "After 'trigger', 'invariant' is valid until either 'release' or 'reaction', and 'reaction' must occur within 'delay' from 'final'." if always trigger = \*, release = \*, delay = \*, final = \*, reaction = \*, invariant = \*. Explain why the resulting sentence is correct.* (1)

This pattern is specified by substituting constant attribute values in various combinations for the "*" symbols. If an attribute is assumed to have a variable value, it is removed from the second part of the prompt. For example, for the combination *release = false*, *delay = true*, *final = true*, *invariant = true* (trigger and reaction are variable), the specific prompt would be written as

*Reformulate in English the following sentence "After 'trigger', 'invariant' is valid until either 'release' or 'reaction', and 'reaction' must occur within 'delay' from 'final'." if always release = false, delay = true, final = true, invariant = true.*

For six attributes and three of their values (*true*, *false*, and *variable*), the number of combinations is quite large: 36 = 729. Therefore, at this stage, it is reasonable to either use automation to substitute all possible attribute combinations or use the results of the pattern's semantic classification, in particular the EDTL [20]. Semantic classification exploits the fact that when constant attribute values are substituted into a pattern, the formal semantics of different combinations of attribute values generate identical formulas. This fact makes it possible to determine canonical combinations of attribute values, taking into account the temporal structure of the resulting formulas. For the EDTL pattern, there are 32 canonical combinations. For example, the combination *release = false*, *delay = true*, *final = true*, *invariant = true* is canonical and produces the semantics $G\ (trigger \rightarrow reaction)$, while the combination *release = false*, *delay = true*, *final = true*, *reaction = true* is not canonical, as it produces the temporally equivalent semantics $G\ (trigger \rightarrow invariant)$.

Prompt (1) is basic and can be used without further additions. However, the following hints to the AI assistant may be useful.

*(2) Formal Semantics of the Requirement.* The formal semantics of the requirement for given values of constant attributes allows the AI assistant to rely not only on the original pattern but also on the semantic formula when constructing the final phrase. The prompt pattern we used for instructions in addition to prompt (1) looks like this:

*The resulting sentence must correspond to the following LTL formula "\*".* (2)

Table 3 shows the results of using only prompt (1) and a combination of prompts (1) and (2).

TABLE III. NL PATTERNS WITH PROMPTS (1) AND (1)+(2)

|   | Attributes | LTL formula | Prompt (1) | Prompt (1)+(2) |
|---|---|---|---|---|
| 1 | rel = false<br>del = true<br>fin = true<br>inv = true | G (trig → rea) | After 'trigger', the condition should be valid until 'rea', which must occur within the specified time limit. | After 'trigger', 'reaction' must occur. |
| 2 | fin = true | G (trig → ((inv ∧ (¬ del)) W (rel v (inv ∧ rea)))) | After 'trigger', 'invariant' should be valid until either 'release' or 'reaction', which must occur within 'delay'. | After 'trigger', 'invariant' must hold and 'delay' must not occur until either 'release' or 'reaction' occurs. |
| 3 | rel = false<br>del = true<br>rea = true | G (trig → ((inv ∧ (¬ fin)) W (fin ∧ inv))) | After 'trigger', 'invariant' should be valid until a point that must occur within a timeframe from 'final'. | After 'trigger', 'invariant' must hold until 'final' occurs. |

Here, in line 1, it is clear that only with prompt (1) does the AI assistant "understand" that constant attributes should not appear in the resulting phrase, but instead simply uses their equivalents: for example, instead of 'invariant' it uses 'condition', instead of 'final' and 'delay' it uses 'time limit'. Such a substitution does not correspond to the meaning of the EDTL requirement. After adding the formula to the prompt, the semantics of the resulting phrase becomes correct. However, the disadvantage of using prompt (2) is that the AI assistant stops paying attention to the original natural language phrase and simply "translates" the LTL formula into natural language, as can be seen from line 2. This leads to cumbersome, unreadable grammatical constructions when there are many non-constant attributes. Therefore, the next experiment excludes prompt (2) but adds explanations for the constant attributes.

*(3) Attribute Name Usage Restrictions.* With constant values for some attributes (in the case of EDTL, these are delay and final), the AI assistant makes incorrect inferences and constructs semantically incorrect statements. For example, if *delay = false*, it assumes the time limit is invalid and there will never be a response (this is clear from the explanations we request from the assistant). In such cases, it is useful to clarify the meaning of the constant attribute values, and we have formulated a prompt of the following form.

*Remember that if invariant is true, then the statement does not depend on the invariant, final = true means that final is now, final = false means that final never happens, delay = false means that the delay is infinite, delay = true means that there is no delay, reaction = false means that we do not wait for the reaction, reaction = true means that the statement does not depend on the reaction.* (3)

Table 4 below shows the results of using combinations of prompt (1) with prompts (2) and (3).

TABLE IV. NL PATTERNS WITH PROMPTS (1)+(2) AND (1)+(3)

|   | Attributes | LTL formula | Prompt (1)+(2) | Prompt (1)+(3) |
|---|---|---|---|---|
| 1 | rel = false<br>del = true<br>fin = true<br>inv = true | G (trig → rea) | After 'trigger', 'reaction' must occur. | After 'trigger', 'reaction' occurs now. |
| 2 | fin = true | G (trig → ((inv ∧ (¬ del)) W (rel v (inv ∧ rea)))) | After 'trigger', 'invariant' must hold and 'delay' must not occur until either 'release' or 'reaction' occurs. | After 'trigger', 'invariant' is valid until either 'release' or 'reaction', and 'reaction' occurs within 'delay' from now. |
| 3 | rel = false<br>del = true<br>rea = true | G (trig → ((inv ∧ (¬ fin)) W (fin ∧ inv))) | After 'trigger', 'invariant' must hold until 'final' occurs | After 'trigger', 'invariant' is valid forever. |

We see, when using prompts (1) and (3), the AI assistant in most cases constructs semantically correct and syntactically readable, uncluttered phrases. However, there are certain combinations of attribute values (in particular, with a constant reaction value) where the resulting phrase has a broader semantics. This occurs because the intuitive meaning of such a combination of attribute values becomes unclear and can only be understood based on the semantic formula. We concluded that such EDTL requirements would also be unclear to humans, so including the corresponding patterns in the CNL makes no sense.

In summary, the combination of prompts (1) and (3) yields better results for constructing a CNL grammar, both in terms of syntax and formal semantics. This combination often allows one to identify combinations of requirement attribute values that do not correspond to a human's intuitive understanding of attribute values. This eliminates the need to use these combinations when specifying requirements in natural language, and instead alerts the engineer to a potential error when specifying requirements in a table form.

*C. Stage 3: CNL Grammar for EDTL*

We present a partial grammar of controlled natural language for EDTL, based on the list of phrase patterns obtained in the previous stage using prompts (1)+(3).

**Req := After <trigger>, <body_trig> | <invariant> is valid <body_inv>**

**<body_trig> := <reaction> occurs <cond_rea> | <invariant> is valid until <cond_inv> | …**

**<cond_rea> := now | from <final> | within <delay> from <final> | …**

## V. Conclusion

We have developed a method for systematically constructing controlled natural language (CNL) requirements based on formal specification patterns using an AI assistant. This method consists of three stages. The first stage involves defining a basic natural language requirement pattern that utilizes all the attributes of a formal specification pattern. The second stage involves using the AI assistant to construct derived natural language requirement patterns with certain attributes. This stage utilizes the formal semantics of the requirement patterns. In the third stage, using the resulting patterns, we construct a CNL grammar. Note that the analysis of the grammatical structure of the patterns is simplified by the LLM generation method used.

Automating the second and third stages will expand the capabilities of CNL construction for arbitrary formal requirement specification patterns.

## Acknowledgment

This work was supported by the Russian Ministry of Education and Science, project 125022803031-1..